\documentstyle[12pt]{article}    
\begin{document}
\begin{center}
   
        {\bf  Microcanonical statistics of black holes and bootstrap
condition\\}

\vspace{1cm}

                      Wung-Hong Huang\\
                       Department of Physics\\
                       National Cheng Kung University\\
                       Tainan,70101,Taiwan\\

\end{center}
\vspace{1cm}
{\bf Abstract}

    The microcanonical statistics of the  Schwarzschild black holes as well
as the Reissner-Nordstr$\sf \ddot{o}$m black holes are analyzed.    In both
cases we set up  the inequalities  in the microcanonical density of states.
   These are then used to show that  the most probable configuration in the
gases of black holes is that one black hole acquires all of the mass and
all of the charge at high energy limit.     Thus the  black holes obey the
statistical bootstrap condition and, in contrast to the other
investigation, we see that U(1) charge does not break the bootstrap
property.   

\vspace{3cm}

      Classification Number: 97.60.Lf, 04.20.Cv

      E-mail:  whhwung@mail.ncku.edu.tw

      Typesetting by Latex

\newpage
\begin{center}  {\bf 1. INTRODUCTION} \end{center}

   The thermodynamics and statistical mechanics of black hole are the
interesting areas of black hole physics [1].    As the black hole may
radiate away completely the incoming pure states will totally evolve into
the outcoming mixed states.    Thus the quantum coherence is lost in the
black hole decay and the unitary principle in the law of quantum mechanics
is violated [2].  

   An attempt to resolve this problem is to take into account the effect of
quantum hair [3,4].    The quantum hair can have dramatic and computable
effects on the thermodynamical behavior of a black hole.   It is hoped that
black holes will carry a lot of quantum hairs which could generate enough
effects to recover the quantum coherence.  

    Another approach, which is more fundamental,  is to find a consistent
theory of quantum gravity which can correctly describe the black hole
radiation.    The only candidate of the quantum gravity, to this day,  is
the string theory [5].    $^,$t Hooft, in a series of inspiring articles,
had shown that black hole behaves like as a special case of string [6].   
Thus the black holes may be thought of being made of string, in some
senses.   Since  the strings carry a lot of massive excitations, thus a lot
of quantum hairs therein may recover the quantum coherence.  The
microcanonical analysis [7] had found that strings obey the bootstrap
condition [8,9]

    $$ \frac {\Omega (E)}{\rho(E)} \to 1, ~~~as~~ E \to \infty.
\eqno{(1.1)}$$
where $\Omega (E)$ is the microcanonical density of states and  $\rho(E)$
the degeneracy of string states.    It is therefore interesting to see
whether the black hole systems obey the bootstrap condition.   

     Historically, the bootstrap model was applied to the statistical model
of the hadrons, in which the hadrons are assumed to be compounds of
hadrons.     The       
bootstrap model of hadron developed by Hagedron [8], Frautschi and others
[9] in the early 70th can be used to explain the ever-increasing number
resonances found in the higher energy.    The statistical mechanics of
black holes suggested by Harms and Leblanc [10] treated the black hole as
composite objects which are made of other black holes, in the spirit of the
old bootstrap model of hadron.    Such a description of black hole might be
of some interesting - in view of the present uncertainty concerning the
black hole entropy problem [11].   In this paper we will reinvestigate the
microcanonical density of states of the  Schwarzschild as well as the
Reissner-Nordstr$\sf\ddot{o}$m black holes, which had been studied by Harms
and Leblanc[10].

   Two kinds of the degeneracy of the states for Schwarzschild  hole will
be considered in section II.   In the natural units ($\hbar = c = G = 1$),
the first form is [4]     

  $$\rho(m) = C exp(4 \pi m^2),\eqno{(1.2)}$$
where m is the mass of the black hole and C is a model-dependent constant. 
The second form is the hole with a quantized spectrum and [12-14]

   $$\rho(n) = C exp(4 \pi n), ~~~ n = 1, 2, ...\eqno{(1.3)}$$  		      
In section III, two kinds of the degeneracy of the states for the
Reissner-Nordstr$\sf\ddot{o}$m hole will be considered.  The first form is
[14]     

    $$\rho(m,q) = C exp[ \pi (m + \sqrt {m^2 - q^2}~)^2],\eqno{(1.4)}$$  		
		      
where m and q are the mass and charge of the hole, respectively.    The
second form is the hole with quantized spectrum and [15]
         
  $$\rho(n,k) = C exp[ \pi (\sqrt{n} + \sqrt {n - k}~)^2], ~~~ k , n = 1,
2, ....(k \le n). \eqno{(1.5)}$$  
Note that the hole mass is quantized in terms of the Plank mass $\sqrt{c
\hbar /G}$ while the electric charge is quantized in terms of the Plank
charge $e/ \sqrt{\alpha}$.

     In both cases we will set up  the useful inequalities  in the
microcanonical density of states.    These are then used to show that  the
most probable configuration in the gases of Schwarzschild is that one black
hole acquires all of the mass in high energy limit.    Thus the
Schwarzschild black holes naturally obey  the bootstrap property.     In
the Reissner-Nordstr$\sf\ddot{o}$m black holes system it shows that  the
most probable configuration is that one black hole acquires all of the mass
and all of the charge at high energy limit.   Thus charged  black holes
obey the statistical bootstrap condition and, in contrast to the other
investigation [10], we see that U(1) charge does not break the bootstrap
property.

\begin{center}  {\bf 2. SCHWARZSCHILD BLACK HOLES} \end{center}
\begin{center}{\bf A. Black holes with continuous spectrum}\end{center}

    The microcanonical density of a gas of Schwarzchild black hole with
continuous spectrum can be written as [9]
 
        $$ \Omega (E,V) = \sum_{N=1}^{\infty} \Omega _N(E,V).\eqno{(2.1)}$$
 The microcanonical density for the configuration with N black holes is
[9,10] 

           $$ \Omega_N (E,V) = \frac{1}{N!} [\frac{V}{(2\pi)^3}]^N
\prod_{i=1}^{N} \int_{m_0}^\infty dm_i \rho (m_i)
\int_{-\infty}^{\infty}dp_i^3  \delta (E-\sum_i E_i) \delta ^3 (\sum_i {\bf
p}_i),             \eqno{(2.2)}$$
where $m_0$ is the lightest mass of the black hole, if it exists.   The
above equation for the density of state first presented by Frautschi (the
Eq.(1.8) in reference 9) was used to investigate the statistical bootstrap
model of hadrons.     It was then adopted by Harms and Leblanc [10] to
investigate the statistical mechanics of black hole, by regarding the black
hole as the compound of black holes. 

  As in [10] we assume that  the black holes obey the dispersion relation,
$m_i^2 = E_i^2 - {\bf p}_i^2$.   Then from Eq.(1.2) we see that  since
$\rho(m_i) = C exp[4 \pi (E_i^2 - {\bf p}_i^2)]$   the high-momentum state
in Eq.(2.2) will contribute negligibly to the momentum integration.  
Therefore we can neglect the momentum-conservation $\delta$ function and 
Eq.(2.2) simply becomes $N$ decoupled Gaussian integrals [9,10]
              $$ \Omega_N (E,V) = \frac{1}{N!} [\frac{V}{(2\pi)^3}]^N
\prod_{i=1}^{N} \int_{m_0}^\infty dE_i \rho (E_i) \delta (E-\sum_i E_i).   
  \eqno{(2.3)}$$

     We now analyze the microcanonical density Eq.(2.3).    When  N=1 then
  
 $$ \Omega_1 (E,V) = \frac{CV}{(2\pi)^3} e^{4\pi E^2} .\hspace{8.5cm}
\eqno{(2.4)} $$   
When  N=2 then

     $$ \Omega_2 (E,V) = \frac{1}{2} ~ [\frac{CV}{(2\pi)^3}]^2  
\int_{m_0}^\infty dE_1 \int_{m_0}^\infty dE_2 e^{4 \pi (E_1^2 + E_2^2)}
\delta (E- E_1-E_2) \hspace{1.5cm}$$
      $$< \frac{1}{2} ~ [\frac{CV}{(2\pi)^3}]^2 \int_0^\infty dE_1
\int_0^\infty dE_2 e^{4 \pi (E_1^2 + E_2^2)} \delta (E- E_1-E_2)$$
      $$  = \frac{1}{2} ~ [\frac{CV}{(2\pi)^3}]^2 \int_0^E dx e^{4 \pi [x^2
+ (E-x)^2]} \hspace{4cm}$$
       $$ = \frac{1}{2} ~ [\frac{CV}{(2\pi)^3}]^2 e^{4 \pi E^2} F(E),
\hspace{5.5cm} \eqno{(2.5)}$$
in which we define 
       $$ F(E) \equiv  \int_0^E dx e^{8 \pi (x^2 - Ex)} < \frac{1}{2\pi
E}(1-e^{-2\pi E^2}) < f_0  \approx  0.255 , \hspace{2cm} \eqno{(2.6)}$$
as shown in the appendix A.    Note that the function F(E) is increasing
from zero, at $E=0$, to the maximum value $f_0$, at $ E \approx 0.445$, 
and then approaches to $\frac{1}{2\pi E}$ at large E .

   To proceed, we see that Eq.(2.3) can be expressed as 
        $$ \Omega_N (E,V) < \frac{1}{N!} [\frac{V}{(2\pi)^3}]^N
\prod_{i=1}^{N} \int_0^\infty dE_i \rho (E_i) \delta (E-\sum_i E_i)  
\hspace{6cm}$$ 
        $$ = \frac{1}{N!} [\frac{CV}{(2\pi)^3}]^N e^{4\pi E^2} \int_0^E
dx_1e^{8\pi E^2 (x_1^2 - E x_1)}  
         \int_0^{E-x_1} dx_2e^{8\pi E^2 [x_2^2 - (E-x_1) x_2]}
\hspace{3cm}$$
        $$ \times\cdots \hspace{9.5cm}$$
         $$ \times \int_0^{E-x_1-x_2 -...-x_{N-3}} dx_{N-2}e^{8\pi E^2
[x_{N-2}^2 - (E-x_1-x_2 -...-x_{N-3}) x_{N-2}]} $$
          $$ \times \int_0^{E-x_1-x_2 -...-x_{N-2}} dx_{N-1}e^{8\pi E^2
[x_{N-1}^2 - (E-x_1-x_2 -...-x_{N-2}) x_{N-1}]} $$
          $$ ~~~~~~ = \frac{1}{N!}~ [\frac{CV}{(2\pi)^3}]^N e^{4\pi E^2}
\int_0^E dx_1e^{8\pi E^2 (x_1^2 - E x_1)} \int_0^{E-x_1} dx_2e^{8\pi E^2
[x_2^2 - (E-x_1) x_2]} \hspace{4cm} $$
         $$ \times\cdots \hspace{9.5cm}$$
         $$\times \int_0^{E-x_1-x_2 -...-x_{N-3}} dx_{N-2}e^{8\pi E^2
[x_{N-2}^2 - (E-x_1-x_2 -...-x_{N-3}) x_{N-2}]} $$
          $$\times F(E-x_1-x_2 -...-x_{N-2}) \hspace{5.2cm}$$
         $$~~~~~~ < \frac{1}{N!}~ [\frac{CV}{(2\pi)^3}]^N ~(f_0)^{N-2}~
e^{4\pi E^2}F(E) \hspace{11cm}$$
         $$~~~~~~ < \frac{1}{N!}~ [\frac{CV}{(2\pi)^3}]^N (f_0)^{N-2}
e^{4\pi E^2} \frac{1}{2\pi E}(1-e^{-2\pi E^2}),\hspace{6.6cm}
\eqno{(2.7)}$$
if $N>2$.
   
   Using this inequality we can obtain the relation 
        $$\sum_{N=2}^{\infty} \Omega _N (E,V) <e^{4\pi E^2} \frac{1}{2\pi
E}(1-e^{-2\pi E^2}) ~~ f_0^{-2}~~ exp[\frac{CV}{(2\pi)^3} f_0].  
\eqno{(2.8)}$$ 
Thus, at high energy limit  
        $$ E >> \frac{(2\pi)^2}{CV} f_0^{-2}exp[\frac{CV}{(2\pi)^3} f_0]
,\eqno{(2.9)}$$
then 
        $$ \Omega _1 (E,V) > \sum_{N=2}^{\infty} \Omega _N (E,V),
\eqno{(2.10)}$$   
and the microcanonical density of a gas of black holes can be approximated
as 
       $$\Omega (E,V)\approx \Omega _1 (E,V) = \frac{CV}{(2\pi)^3} e^{4 \pi
E^2}. \eqno{(2.11)}$$ 
Thus the most probable configuration for a gas of Schwarzschild  black
holes with continuous spectrum will be that  at N =1.    This implies that
one hole acquires all of the mass and the  bootstrap condition is obeyed.

\begin{center}  {\bf B. Black holes with discrete spectrum} \end{center}

   Next, we  investigate the black holes system with discrete spectrum.  
The microcanonical density a gas of black holes is written as  
           $$ \Omega (E) = \sum_{N=1}^{\infty} \Omega (N,E).
\eqno{(2.12)}$$
The density for the configuration with N black holes is 
           $$ \Omega (N,E) = \frac {1}{N!} \prod_{i=1}^{N} \sum_{l_i       
         =1}^{\infty}  \rho(l_i) ~ \delta _{E,  \sum_{i =1}^{N}  E_{l_i}} =
\frac {1}{N!} \prod_{i=1}^{N} \sum_{l_{i} =1}^{\infty}   C g^{l_i} ~ \delta
_{E,  \sum_{i =1}^{N}  \sqrt{l_i}}~ ,                        
\eqno{(2.13)}$$
in which $g \equiv e^{4\pi}$ according to the Eq.(1.3).   Note that the
value of $g$ (order 1) which may be model-dependent [12-14] does not affect
the following analysis. 

 We now analyze the microcanonical density Eq.(2.13).    When  N=1 then   
           $$ \Omega (1,E) = C  g^{E^2} . \hspace{4cm}\eqno{(2.14)}$$
When N =2 then 

           $$ \Omega (2,E) = \frac {C^2}{2} \sum_{l_1=1}^{\infty}          
                       \sum_{l_2=1}^{\infty}  g^{l_1} g^{l_2} \delta _{E, 
\sqrt{l_1} +                                \sqrt{l_2}} ~~ $$
           $$=\frac {C^2}{2}   g^{E^2} \sum_{l=1}^{(E-1)^2} g^{2(l-        
      E\sqrt{l})}$$
              
           $$ \equiv \frac {C^2}{2}   g^{E^2} K(E,g) . ~~~~~
\eqno{(2.15)}$$
Now, through a simple calculation we can see that $K(E,e^{4\pi})$ is a
rapidly decaying function with respect to the variable $E$.    For
examples, $K(2,e^{4\pi}) \approx 10^ {-6}, K(5,e^{4\pi}) \approx 10^
{-22}$, ...,$ K(10,e^{4\pi}) \approx 10^ {-49}$ .   Note that the hole mass
is quantized in terms of the "Plank mass".
Keep the property of the function K in mind  we see that Eq.(2.13) can be
expressed as 
          $$ \Omega (N,E) = \frac {C^N}{N!} ~ g^{E^2} ~ \sum_{l_{1}
=1}^{(E-(N-1))^2} g^{2(l_{1}-E\sqrt{l_1})}
 \sum_{l_{2} =1}^{(E-(N-2)- \sqrt{l_1})^2} g^{2(l_{2} - \sqrt{l_2} ( E-
\sqrt{l_1}))}  \hspace{4.5cm} $$
                  $$~~~~ \times \cdots \hspace{9cm} $$
          $$ \times \sum_{l_{N-2} =1}^{(E-2- \sqrt{l_1}- .... -
\sqrt{l_{N-3}})^2} g^{2(l_{N-2} - \sqrt{l_{N-2}} ( E- \sqrt{l_1}- .... -
\sqrt{l_{N-3}}))} $$
         $$ \times \sum_{l_{N-1} =1}^{(E-1- \sqrt{l_1}- .... -
\sqrt{l_{N-2}})^2} g^{2(l_{N-1} - \sqrt{l_{N-1}} ( E- \sqrt{l_1}- .... -
\sqrt{l_{N-2}}))} $$
         $$ < ~~ \frac {C^N}{N!} ~ g^{E^2} ~ \sum_{l_{1} =1}^{(E-1)^2}
g^{2(l_{1}-E\sqrt{l_1})}
 \sum_{l_{2} =1}^{(E- \sqrt{l_1})^2} g^{2(l_{2} - \sqrt{l_2} ( E-
\sqrt{l_1}))}\hspace{6cm}   $$
         $$~~~~ \times \cdots \hspace{9cm} $$
         $$ \times \sum_{l_{N-2} =1}^{(E- \sqrt{l_1}- .... -
\sqrt{l_{N-3}})^2} g^{2(l_{N-2} - \sqrt{l_{N-2}} ( E- \sqrt{l_1}- .... -
\sqrt{l_{N-3}}))} $$
         $$ \times \sum_{l_{N-1} =1}^{(E- \sqrt{l_1}- .... -
\sqrt{l_{N-2}})^2} g^{2(l_{N-1} - \sqrt{l_{N-1}} ( E- \sqrt{l_1}- .... -
\sqrt{l_{N-2}}))} $$
                  $$ < ~~ \frac {C^N}{N!}  g^{E^2} \sum_{l_{1}
=1}^{(E-1)^2} g^{2(l_{1}-E\sqrt{l_1})}  K(1,g)^{N-1} \hspace{8.5cm} $$
         $$ < ~~ \frac {C^N}{N!}  g^{E^2}  K(E,g).\hspace{10.5cm} 
\eqno{(2.16)} $$
if $N>2$.    

   Thus 
         $$ \sum_{n=2}^{\infty} \Omega (N,E) < g^{E^2} (\frac {C^2} {2!} +
\frac {C^3} {3!} + .... ) K(E,g) \hspace{2cm}  $$
         $$  =  g^{E^2}  K(E,g)  (e^C -1).  \eqno{(2.17)}$$
Since the function  $K(E,g)$ is a rapidly decaying function with respect to
the variable $E$ we  conclude that 
         $$ \Omega (N, E) \approx \Omega (1,E) = C g^{E^2} ,
\eqno{(2.18)}$$
if the energy of the system is sufficiently large.    Thus the most
probable configuration for a gas of Schwarzschild  black holes with
quantized spectrum will be that at N =1.    This means that one hole
acquires all of the mass and
the  bootstrap condition is obeyed.    Note that  the term corrected to the
Hawking$^{,}$s temperature found in Ref.[16] in the canonical ensemble of
black holes (for example, the eq.(26) in  [16]) does not show in the 
microcanonical treatment of  this paper.

\begin{center}  {\bf  3. REISSNER-NORDSTR$\sf\ddot{O}$M BLACK HOLES} 
\end{center}

    The analysis of microcanonical density of a gas of the
Reissner-Nordstr$\sf\ddot{o}$m black holes is very similar to that in the
Schwarzchild   black holes.   To begin with, let us mention the main point
in the section II.   In there we first show that the two-hole density 
$\Omega_2 (E,V)$ (or $\Omega (2,E)$) is less then the product of one-hole
density  $\Omega_1 (E,V)$ (or $\Omega (1,E)$) by  an energy-dependent
function $F(E)$ (or $K(E,g)$).    {\it The crucial property is that this
function is never  larger then one and will approach to zero at high
energy.}    This result can also be used  to see that the N-hole density is
always  less then ($N-1$)-hole density.    Then, repeatedly using this
property we thus show that the N-hole density is small then the one-hole
density at high energy limit. 

      Therefore, the only work we now need to do is to show that the 
two-Reissner-Nordstr$\sf\ddot{o}$m-hole density is less then the product of
one-Reissner-Nordstr$\sf\ddot{o}$m-hole density by  an energy-dependent
function, and this function is never larger then one and shall approach to
zero at high energy.    

\begin{center}{\bf A.  Charged black holes with continuous spectrum}
\end{center}

     Let us first analyze the case with continuous spectrum.    The
microcanonical density for the configuration with N charged black holes is
[10] 

           $$ \Omega_N (E,Q,V) = \frac{1}{N!} [\frac{V}{(2\pi)^3}]^N
\prod_{i=1}^{N} \int_{m_0}^\infty dm_i  \int_{- m_i}^{m_i} dq_i
\int_{-\infty}^\infty dp_i^3  \hspace{4cm} $$
$$ \times \rho (m_i,q_i) \delta (E-\sum_i E_i) \delta (Q-\sum_i q_i) \delta
^3 (\sum_i {\bf p}_i),         \eqno{(3.1)}$$
where $m_0$ is the lightest mass of the black hole, if it exists.    Once
again, we assume that  black holes obey the dispersion relation, $m_i^2 =
E_i^2 - {\bf p}_i^2$.   Then, likes that in the neutral holes, the
high-momentum state in Eq.(3.1) will contribute negligibly to the momentum
integration and  we can neglect the momentum-conservation $\delta$
function.   Thus  Eq.(3.1) becomes $N$ decoupled Gaussian integrals [10]

              $$ \Omega_N (E,Q,V) = \frac{1}{N!} [\frac{V}{(2\pi)^3}]^N
\prod_{i=1}^{N} \int_{m_0}^\infty dE_i \int_{-E_i}^{E_i} dq_i ~
\rho(E_i,q_i) \delta (E-\sum_i E_i) \delta (Q-\sum_i q_i).  \eqno{(3.2)}$$

     We now analyze the microcanonical density Eq.(3.2).    When  N=1 then
from Eq.(1.4) we have   
 $$ \Omega_1 (E,Q,V) = C exp[ \pi (E + \sqrt {E^2 - Q^2} ~
)^2].\hspace{6cm} \eqno{(3.3)} $$   
When  N=2 then
     $$ \Omega_2 (E,Q,V) = \frac{1}{2} ~ [\frac{CV}{(2\pi)^3}]^2  
\int_{m_0}^\infty dE_1 \int_{-E_i}^{E_i} dq_1 \int_{m_0}^\infty dE_2
\int_{-E_2}^{E_2} dq_2 \hspace{5cm}$$
        $$\times  exp[ \pi (E_1 + \sqrt {E_1^2 - q_1^2}~)^2 + \pi (E_2 +
\sqrt {E_2^2 - q_2^2}~)^2] $$
$$\times \delta (E- E_1-E_2) \delta (Q- q_1-q_2) \hspace{3cm}$$
  $$~~~~< \frac{1}{2} ~ [\frac{CV}{(2\pi)^3}]^2   \int_0^\infty dE_1
~~2\int_0^{E_i} dq_1 \int_0^\infty dE_2 ~~2 \int_0^{E_2} dq_2
\hspace{3cm}$$
        $$\times  exp[ \pi (E_1 + \sqrt {E_1^2 - q_1^2}~)^2 + \pi (E_2 +
\sqrt {E_2^2 - q_2^2}~)^2] $$
$$\times \delta (E- E_1-E_2) \delta (Q- q_1-q_2) \hspace{3cm}. 
\eqno{(3.4)} $$

\noindent  
Since $E  = E_1 + E_2$  we have the relation   

$$ (E_1 + \sqrt {E_1^2 - q_1^2}~)^2 +  (E_2 + \sqrt {E_2^2 - q_2^2}~)^2
\hspace{8cm}$$
$$ =  (E_1^2 + E_2^2) + (E_1^2 - q_1^2) + (E_2^2  - q_2^2) + 2E~ ( \sqrt
{E_1^2 - q_1^2} +   \sqrt {E_2^2 - q_2^2}~)\hspace{3cm} $$
$$~~~~~~~~~~~~~~~~~~~~~~~~~~~~~~ - 2  (E_2  \sqrt {E_1^2 - q_1^2} +  E_1
\sqrt {E_2^2 - q_2^2}~)$$
$$ <  (E_1^2 + E_2^2) + (E_1^2 - q_1^2) + (E_2^2  - q_2^2) + 2E~ ( \sqrt
{E_1^2 - q_1^2} +   \sqrt {E_2^2 - q_2^2}~) - 2 E_2  \sqrt {E_1^2 - q_1^2}.
  $$
$$\eqno{(3.5)}$$

   To proceed,  using the appendices B and C we see that 

$$(E_1^2 - q_1^2) + (E_2^2  - q_2^2) < E^2 - Q^2, \eqno{(3.6)}$$
$$ 2E~ ( \sqrt {E_1^2 - q_1^2} +   \sqrt {E_2^2 - q_2^2}~) < 2 E \sqrt{E^2
- Q^2},   \eqno{(3.7)}$$
$$\int_0^{E_1} dq_1  \int_0^{E_2} dq_2 exp(-2 \pi E_2  \sqrt {E_1^2 -
q_1^2}) \delta (Q - q_1 - q_2) < H(Q) < h_0 \approx 0.275 . \eqno{(3.8)}$$

\noindent
Through the similar prescription as that in deriving Eq.(2.5) we have the
relation

 $$   \int_0^\infty dE_1 \int_0^\infty dE_2 ~e^{\pi (E_1^2 + E_2^2)}
~\delta (E- E_1-E_2) <  e^{ \pi E^2} G(E),  \eqno{(3.9)}$$

\noindent
in which we define 
       $$ G(E) \equiv  \int_0^E dx e^{2 \pi (x^2 - Ex)} < \frac{2}{\pi
E}(1-e^{- \pi E^2 /2}) < g_0  \approx  0.509 . \hspace{2cm} \eqno{(3.10)}$$
Note that the function $G(E)$ is increasing from zero, at $E=0$, to the
maximum value $g_0$, at $E \approx 0.894$, and then approaches to
$\frac{2}{\pi E}$ at large E .

    Substituting the inequalities Eqs.(3.6)-(3.10) into Eq.(3.4) we thus
have the  inequality  

$$\Omega_2 (E,Q,V) < \frac{1}{2} ~ [\frac{CV}{(2\pi)^3}]^2 ~4h_0 G(E)
exp[\pi (E + \sqrt{E^2 - Q^2}~)^2] .  \eqno{(3.11)}$$

Now we have shown that  the two-hole density $ \Omega_2 (E,Q,V)$  is less
then the product of one-hole density $ \Omega_1 (E,Q,V)$ by  an
energy-dependent function $G(E)$ which  is never larger then one and will
approach to zero at high energy.     Then, as that in section II, after
repeatedly using this property we can easily find that  

 $$\sum_{N=2}^{\infty} \Omega _N (E,Q,V) <e^{\pi E^2} \frac{2}{\pi
E}(1-e^{-\pi E^2 /2}) ~~ (4 h_0 f_0)^{-2}~~ exp[\frac{CV}{(2\pi)^3}(4 h_0
f_0)].   \eqno{(3.12)}$$ 

\noindent
Therefore at high energy limit  
        $$ E >> \frac{(2\pi)^2}{CV} (4 h_0 f_0)^{-2}exp[\frac{CV}{(2\pi)^3}
(4 h_0 f_0)] ,\eqno{(3.13)}$$
then 
        $$ \Omega _1 (E,Q,V) > \sum_{N=2}^{\infty} \Omega _N (E,Q,V),
\eqno{(3.14)}$$   
and the microcanonical density of a gas of charged black holes can be
approximated as 
       $$\Omega (E,Q,V)\approx \Omega _1 (E,Q,V) =\frac{CV}{(2\pi)^3} exp[
\pi (E + \sqrt {E^2 - Q^2} ~ )^2]. \eqno{(3.15)}$$

Thus the most probable configuration for a gas of 
Reissner-Nordstr$\sf\ddot{o}$m black holes with continuous spectrum will be
that  at N =1.    This implies that one hole acquires all of the mass and
all of the charge.    Note that the  bootstrap condition is obeyed in our
analysis and thus the U(1) charge does not break the bootstrap property of
the black hole, in contrast to the claim in [10].

\begin{center}  {\bf B. Charged black holes with discrete spectrum}
\end{center}

   Next, we  investigate the charged black holes system with discrete
spectrum.   The microcanonical density  is written as  

           $$ \Omega (E,Q) = \sum_{N=1}^{\infty} \Omega (N,E,Q).
\eqno{(3.16)}$$

\noindent
The density for the configuration with N charged black holes is 
           $$ \Omega (N,E,Q) = \frac {1}{N!} \prod_{i=1}^{N} \sum_{n_i     
           =1}^{\infty} \sum_{k_i=1}^n  \rho(n_i, k_i) ~ \delta _{E, 
\sum_{i =1}^N \sqrt {n_i}} \delta _{Q,  \sum_{i =1}^N  \sqrt {k_i}}
~~~~\hspace{5cm} $$
           $$ = \frac {1}{N!} \prod_{i=1}^{N} \sum_{n_i =1}^{\infty}
\sum_{k_i=1}^{n_i}   C exp[ \pi (\sqrt{n_i} + \sqrt {n_i - k_i}~)^2] ~
\delta _{E,  \sum_{i =1}^N \sqrt {n_i}} \delta _{Q,  \sum_{i =1}^N  \sqrt
{k_i}} \eqno{(3.17)}$$

 When N=1 then   
         $$ \Omega (1,E,Q) = C exp[ \pi (E + \sqrt {E^2 - Q^2}~)^2] .
\hspace{4cm}\eqno{(3.18)}$$
When N =2 then 
$$ \Omega (2,E,Q) = \frac {C^2}{2} \sum_{n_1 =1}^{\infty}
\sum_{k_1=1}^{n_1} \sum_{n_2 =1}^{\infty} \sum_{k_2=1}^{n_2}  exp[ \pi
(\sqrt{n_1} + \sqrt {n_1 - k_1}~)^2] exp[ \pi (\sqrt{n_2} + \sqrt {n_2 -
k_2}~)^2] $$
$$\times ~ \delta _{E,  \sqrt {n_1}+\sqrt{n_2}} ~\delta _{Q,  \sqrt
{k_1}+\sqrt{k_2}}. \eqno{(3.19)}$$

Since $E  = \sqrt{n_1} + \sqrt{n_2}$  we have the relation   
$$  (\sqrt{n_1} + \sqrt {n_1 - k_1}~)^2  +  (\sqrt{n_2} + \sqrt {n_2 -
k_2}~)^2   \hspace{11cm}$$
$$ = (n_1 + n_2) + (n_1 - k_1) +(n_2 - k_2)  + 2E~ (\sqrt {n_1} \sqrt {n_1
- k_1} + \sqrt {n_2} \sqrt {n_2 - k_2} ~)\hspace{2.5cm} $$
$$~~~~~~~~~~~~~~~~~~~~~~~~~~~~~~~~~~~~~~~~ - 2  (\sqrt {n_2} \sqrt {n_1 -
k_1} + \sqrt {n_1} \sqrt {n_2 - k_2}~)\hspace{2cm}$$
$$ <  (n_1 + n_2) + (n_1 - k_1+ n_2 - k_2) + 2E~ (\sqrt {n_1} \sqrt {n_1 -
k_1} + \sqrt {n_2} \sqrt {n_2 - k_2} ~) - 2 \sqrt {n_2} \sqrt {n_1 - k_1}. 
\eqno{(3.20)}$$

   Using the identities shown in the appendices B and C we have  

$$(n_1 - k_1 + n_2  - k_2) < E^2 - Q^2, \eqno{(3.21)}$$
$$ 2E~ ( \sqrt {n_1 - k_1} +  \sqrt {n_2 - k_2}~) < 2 E \sqrt{E^2 - Q^2},  
\eqno{(3.22)}$$
$$ \sum_{k_1=1}^{n_1} \sum_{k_2=1}^{n_2}       exp[ - 2\pi  \sqrt {n_2}
\sqrt {n_1 - k_1}]  \delta _{Q,  \sqrt {k_1}+\sqrt{k_2}}< J(Q) \le 1 .
\eqno{(3.23)}$$

\noindent
Using the similar prescription as that in deriving Eq.(2.15) we have the
relation

$$\sum_{n_1=1}^\infty \sum_{n_2=1}^\infty exp[ \pi (n_1 +n_2)]  \delta _{E,
 \sqrt {n_1}+\sqrt{n_2}}<   K(E,e^\pi) exp(\pi E^2) ,   \eqno{(3.24)}$$ 

\noindent
in which, as mentioned in Eq.(2.15), $K(E,e^\pi)$ is a rapidly decaying
function with respect to the variable $E$.    For examples, $K(2,e^\pi)
\approx 0.04, K(5, e^\pi) \approx 10^{-6}, \cdots , K(10,e^\pi) \approx
10^{-12}.$    Using the inequalities Eqs.(3.21)-(3.24) we thus have the
inequality  

$$\Omega_2 (E,Q,V) < \frac{C^2}{2} ~K(E,e^\pi) exp[\pi (E + \sqrt{E^2 -
Q^2}~)^2] .  \eqno{(3.25)}$$

    As we have shown that  two-hole density $ \Omega (2,E,Q)$  is less then
the product of one-hole density $ \Omega (1,E,Q)$ by  an energy-dependent
function $K(E,e^\pi)$ which is never larger then one and approaches to zero
at high energy.     Then, as that in section II, repeatedly using this
property we can easily find that  

        $$ \sum_{n=2}^{\infty} \Omega (N,E,Q) <  K(E,e^\pi)  (e^C -1)
exp[\pi (E + \sqrt{E^2 - Q^2}~)^2].  \eqno{(3.26)}$$
Since the function  $K(E,e^\pi)$ is a rapidly decaying function with
respect to the variable $E$ we  conclude that 
         $$ \Omega (N, E,Q) \approx \Omega (1,E,Q) = C  exp[\pi (E +
\sqrt{E^2 - Q^2}~)^2] , \eqno{(3.27)}$$
if the energy of the system is sufficiently large.    Thus the most
probable configuration for a gas of  Reissner-Nordstr$\sf\ddot{o}$m  black
holes with quantized spectrum will be that at N =1.    This means that one
hole acquires all of the mass and all of charge.    Thus the  bootstrap
condition is still obeyed.

\begin{center}  {\bf 4. CONCLUSION} \end{center}

    In conclusion,   I  have used the microcanonical treatment to study the
statistical mechanics of a gas of Schwarzschild black holes or
Reissner-Nordstr$\sf\ddot{o}$m black holes.    The black holes may have the
discrete spectrum or have the continuous spectrum .    In these systems  I 
have set up  the inequalities in  the microcanonical ensemble of N black
holes.    The central ideal in our treatment is first to show that  the
two-hole density is always less then the product of one-hole density by an
energy-dependent function.    This function is found to be never  larger
then one and will approach to zero at high energy.    Once this relation is
established then it can be adopted to show that the $N$-hole density is
always  less then the ($N-1$)-hole density.    Then, repeatedly using this
property we thus finally show that the N-hole density is small then the
one-hole density at high energy limit.    Thus the most probable
configuration is that with N =1, if the energy of the system is
sufficiently large.   This implies that the bootstrap condition is obeyed
in the black holes system and  U(1) charge does not break the bootstrap
property.

\newpage

\noindent
 {\bf APPENDIX A}

       From the definition 
  $$ F(E) \equiv \int_0^E dx e^{8 \pi (x^2 - Ex)} =  E \int_0^1 dy e^{8 \pi
E^2(y^2 - y)} \hspace{4cm}$$
       $$ = E e^{-2\pi E^2} \int_0^1 dy e^{8\pi E^2 (y-1/2)^2}= 2E e^{-2\pi
E^2} \int_0^{1/2} dz e^{8\pi E^2 z^2} $$
       $$ <2 E e^{-2\pi E^2} \int_0^{1/2} dz e^{4\pi E^2 z} = \frac{1}{2\pi
E}(1-e^{-2\pi E^2}),  \hspace{1.5cm}   \eqno{(A1)}               $$
which is used in Eq.(2.5)\\

\noindent
{\bf APPENDIX B} 

    From the figure  1 we see that  
$$(E_1^2 - q_1^2) + (E_2^2  - q_2^2) \equiv  \overline{ac}^2 -
\overline{ab}^2 + \overline{cd}^2 - \overline{ce}^2 = \overline{cb}^2 +
\overline{de}^2 \hspace{4cm} $$ 
$$ ~~~~~~~~~= \overline{ef}^2 + \overline{de}^2 <  (\overline{ef} +
\overline{de})^2 = \overline{ad}^2 - \overline{abf}^2  $$
$$ < (\overline{ac} + \overline{cd})^2 - \overline{abf}^2  = E^2 - Q^2 ,
\eqno{(B1)}$$ 
which is used in Eq.(3.6).

    From figure 1 we also see that  
$$  \sqrt {E_1^2 - q_1^2} +  \sqrt {E_2^2 - q_2^2} \equiv 
\sqrt{\overline{ac}^2 - \overline{ab}^2} + \sqrt{\overline{cd}^2 -
\overline{ce}^2} = \overline{cb} + \overline{de} \hspace{4cm} $$ 
$$ ~~~~~~~~~= \overline{ef} + \overline{de} = \overline{def} = 
\sqrt{\overline{ad}^2 - \overline{abf}^2} ~~~~~~~~~~~~~~~ $$
$$ < \sqrt{(\overline{ac} + \overline{cd})^2 - \overline{abf}^2}  =
\sqrt{E^2 - Q^2} ,\eqno{(B2)}$$
which is used in  Eq.(3.7).

   Once letting $n_1= E_1^2, n_2 = E_2^2, k_1= q_1^2 $ and $k_2 = q_2^2$,
then the above inequalities also imply the relations used in Eqs.( 3.21)
and (3.22).\\

\noindent
{\bf APPENDIX C} 

   Since $q_2 \leq E_2$  we have 
   
$$\int_0^{E_1} dq_1  \int_0^{E_2} dq_2 exp(-2 \pi E_2  \sqrt {E_1^2 -
q_1^2}) \delta (Q - q_1 - q_2)~~~~~~~~~$$
$$ ~~~~~~~~~~~~~~~~~~~\le \int_0^{E_1} dq_1  \int_0^{E_2} dq_2 exp(-2 \pi
q_2  \sqrt {E_1^2 - q_1^2}) \delta (Q - q_1 - q_2)$$
$$ =  \int_0^{E_1} dq_1 exp[-2 \pi (Q - q_1) \sqrt {E_1^2 - q_1^2}~].
\eqno{(C1)}$$

    (i) When $Q \geq E_1$ (note that $E = E_1 + E_2 \geq Q$) then Eq.(C1)
becomes

$$  \int_0^{E_1} dq_1 exp[-2 \pi (Q - q_1) \sqrt {E_1^2 - q_1^2}~] \leq  
\int_0^{E_1} dq_1 exp[-2 \pi (E_1 - q_1) \sqrt {E_1^2 - q_1^2}~] \equiv
H(E_1). $$
$$\eqno{(C2)}$$
After a simple calculation we see that the function H is increasing from
zero, at $E_1 = 0$,  to the maximum value  $h_0 \approx 0.275$, at $E_1
\approx 0.506$,  and then rapidly approaches to zero at large $E_1$. 

   (ii) When $Q < E_1$ then since $ 0 \leq q_2 =Q -q_1$ we have to
constrain the integration of $q_1$ in Eq.(C1) to be from zero to $Q$, thus 
 
$$  \int_0^{E_1} dq_1 exp[-2 \pi (Q - q_1) \sqrt {E_1^2 - q_1^2}~] \leq  
\int_0^{Q} dq_1 exp[-2 \pi (Q - q_1) \sqrt {Q^2 - q_1^2}~] \equiv H(Q).
\eqno{(C3)}$$
As before,  $H(Q) < h_0 \approx 0.275$.     This establishes the inequality
Eq.(3.8)    

   The prove the relation Eq.(3.23) we can let $n_1= E_1^2, n_2 = E_2^2,
k_1= q_1^2 $ and $k_2 = q_2^2$ in the above treatment.     Then we have the
relation 
$$ \sum_{k_1=1}^{n_1} \sum_{k_2=1}^{n_2}       exp[ - 2\pi  \sqrt {n_2}
\sqrt {n_1 - k_1}]  \delta _{Q,  \sqrt {k_1}+\sqrt{k_2}} < 
\sum_{k_1=1}^{Q^2}  exp[ - 2\pi  (Q - \sqrt {k_1}~) \sqrt {Q^2 - k_1}]
=J(Q). \eqno{(C4)}$$
After a simple calculation we see that the function J(Q) is a rapidly
decaying function with respect to the variable $Q$.       For examples,
$J(1) =1, J(5) \approx 10^{-34}, \cdots , J(10) \approx 10^{-78}$.

\newpage

\begin{enumerate}

\item  J. D. Bekenstein, Phys. Rev. D 7 (1973) 2333; S. W. Hawking, Commun,
                                       Phys. 43 (1975) 199; Phys. Rev. D 13
(1976) 191.
\item  S. W. Hawking, Phys. Rev. D 14 (1976) 2460.
\item  L. M. Krauss and F. Wilczek, Phys. Rev. Lett. 62 (1989) 1221.
\item  S. Coleman, J. Preskill, and F. Wilczek, Nucl. Phys. B378, 175
(1992).
\item  M. Green, J. Schwarz and E. Witten, Superstring Theory (Cambridge   
              University Press, Cambridge, 1987).
\item   G. $^,$t Hooft,  Nucl. Phys. B335, 138 (1990), and references
therein.
\item  M. Bowick and L. Wijewardhana, Phys. Rev. Lett. 54, 2485 (1955).
\item R. Hagedorn, Nuove Cimento Suppl. 3, 147 (1965).
\item S. Frautschi, Phys. Rev. D 3, 2821 (1971); R. D. Carlitz, Phys. Rev.
D 5, 3231 (1972).
\item B. Harms and Y. Leblanc, Phys. Rev. D 46, 2334 (1992); D 47, 2438    
                                    (1993).
\item V. P. Prolov and D. V. Fursaev, Class. Quantum Grav. 15, 2041 (1998).
\item J.  D. Bekenstein ,   Lett. Nuovo Cimento  11 (1974) 467; J.  D.
Bekenstein and V. F. Mukhanov,  Phys. Lett. B 360 (1995) 7.
\item C. O. Lousto, Phys. Rev. D 51 (1995) 1733; C. O. Lousto and J.
Makela,  Phys. Rev. D 54 (1996) 4982.
\item T. Brotz and C. Kiefer,  Phys. Rev. D 55 (1997) 2186; V. A. Berezin,
Phys. Rev. D 55 (1997) 2139; C. Vaz and L.Witten, Phys. Rev. D 60 (1999)
024009; Y. Peleg, Phys. Lett. B 356 (1995) 462; J. Makela, Phys. Lett. B
390 (1997) 115;  H. A. Kastrup,  Phys. Lett. B 385 (1996) 75.   
\item J. Makela and P. Repo,  Phys. Rev. D 57 (1998) 4899.
\item H. A. Kastrup,  Phys. Lett. B 413 (1997) 267.

\end{enumerate}
\newpage
\unitlength 2mm
\begin {picture}(40,50)
\put(0,0){\line(3,1){30}}
\put(30,10){\line(1,2){10}}
\put(0,0){\line(4,3){40}}
\put(0,0){\line(1,0){40}}
\put(30,10){\line(1,0){10}}
\put(30,0){\line(0,1){10}}
\put(40,0){\line(0,1){30}}
\put(0,2){d}
\put(31,1){e}
\put(41,0){f}
\put(29,11){c}
\put(41,10){b}
\put(41,30){a}
\end {picture}
\\

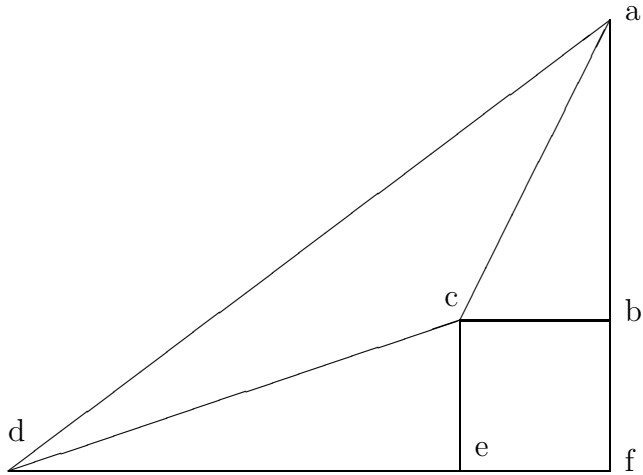
   FIG.1 Since $E_1 \ge q_1$ and  $E_2 \ge q_2$ we can let $\bar{ac} = E_1,
\bar{ab} = q_1, \bar{cd} = E_2$ and $\bar{ce} = q_2$ .  Note that $\angle
abc = \angle ced = \pi /2$.
\end{document}